# Improving the Load Balance of MapReduce Operations based on the Key Distribution of Pairs

Liya Fan[1], Bo Gao[1], Xi Sun[1], Fa Zhang[2], Zhiyong Liu[2]
1 IBM China Research Laboratory
2 Institute of Computing Technology, Chinese Academy of Sciences
Beijing China
fanliya@cn.ibm.com, tju_gb@163.com, alvinswen@gmail.com, fazhang@gmail.com, zyliu@ict.ac.cn

*Abstract*—Load balance is important for MapReduce to reduce job duration, increase parallel efficiency, etc. Previous work focuses on coarse-grained scheduling. This study concerns fine-grained scheduling on MapReduce operations. Each operation represents one invocation of the Map or Reduce function. Scheduling MapReduce operations is difficult due to highly skewed operation loads, no support to collect workload statistics, and high complexity of the scheduling problem. So current implementations adopt simple strategies, leading to poor load balance. To address these difficulties, we design an algorithm to schedule operations based on the key distribution of intermediate pairs. The algorithm involves a sub-program for selecting operations for task slots, and we name it the Balanced Subset Sum (BSS) problem. We discuss properties of BSS and design exact and approximation algorithms for it. To transparently incorporate these algorithms into MapReduce, we design a communication mechanism to collect statistics, and a pipeline within Reduce tasks to increase resource utilization. To the best of our knowledge, this is the first work on scheduling MapReduce workload at this fine-grained level. Experiments on PUMA [T+12] benchmarks show consistent performance improvement. The job duration can be reduced by up to 37%, compared with standard MapReduce.

## 1 Introduction

MapReduce has emerged as a powerful computing framework for processing big data in Cloud and distributed computing [DG04]. It has some distinct advantages over traditional frameworks, like MPI [Pa96] and PVM [G+96]. For example, users of MapReduce do not have to care about messy details like data distribution, fault tolerance, load balance, etc. When a job is submitted, the framework will take care of these issues automatically. Due to its robustness and ease of use, MapReduce has gained popularity in both research and industry forums.

MapReduce performs two basic functions: Map and Reduce. It works by partitioning the workload of a job into a set of Map/Reduce tasks. Each Map/Reduce task is further divided into one or more Map/Reduce operations. In particular, each invocation of the Map/Reduce function is named a Map/Reduce *operation*. These operations are distributed across available Map/Reduce task slots and executed in parallel. Therefore, achieving load balance for MapReduce is an important problem. It determines the parallel efficiency, resource utilization, job duration, etc. Previous work focuses on coarse-grained scheduling of MapReduce workload. For example, [K+11] [SL10] [B+05] and [BD11] address the load balance problem at the job or task level. In this study, we consider the problem at a fine-grained level: the operation level.

We define the *load* of a task or operation as the number of key-value pairs to be processed by the task or operation. We will show that achieving load balance for MapReduce operations is difficult, especially for Reduce operations. Several new challenges arise: 1) The loads of operations are highly skewed [K+11b]. For example, Fig. 1(a) shows the 80 Reduce operation loads generated by a benchmark of PUMA (Purdue MapReduce benchmarks suite) [T+12]. Such skew will easily cause load imbalance of task slots if not managed properly, and the performance loss is notable. 2) To achieve load balance, the load of each operation should be obtained [G+12]. This depends on the key distribution of intermediate pairs. However, it is difficult to obtain such distribution by the current MapReduce specification. 3) Even if the loads of all operations are known, the load balance problem is difficult. We will show that the problem is equivalent to a strongly NP-hard problem [Ho98]. Due to these difficulties, existing MapReduce implementations make simple assumptions about operations, and adopt simple scheduling strategies [DG04], resulting in poor load balance. Fig. 1(b) illustrates the loads for task slots produced by standard MapReduce on the same benchmark as Fig. 1(a). It can be seen that the largest load is 673 times greater than the smallest load.

To address the above difficulties, we devise a communication mechanism to extend the current MapReduce specification. It collects the key distribution of intermediate pairs. Based on the distribution, we model the load balance problem as a dynamic program, and show it is equivalent to a scheduling problem denoted as P||$C_{max}$ [Ho98]. To solve this problem, we introduce a novel algorithm design paradigm named *dynamic programming decomposition*. Through this technique, the scheduling problem is reduced to a series of simpler problem instances. We name this simpler problem Balanced Subset Sum (BSS) problem. The BSS problem is weakly NP-hard. We give exact and approximation algorithms for it. All these algorithms and mechanism have been implemented and transparently incorporated into MapReduce.

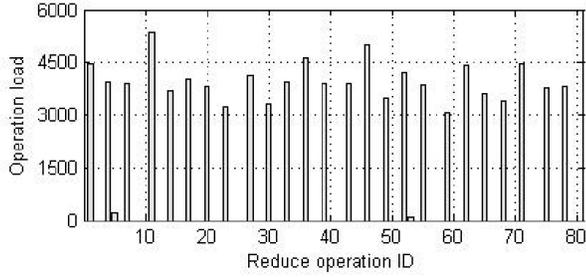

(a) Reduce operations loads

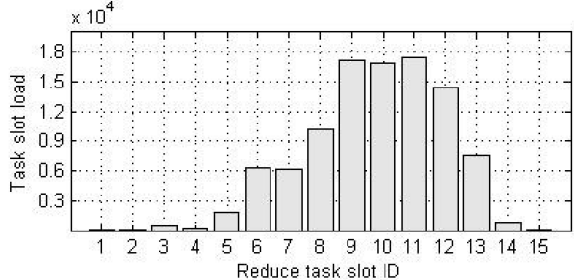

(b) Reduce task slot loads produced by standard MapReduce
Figure 1. Skews produced by standard MapReduce. The benchmark is Histogram-Movies from PUMA. The input data is from PUMA, of 5.1 GB. The MapReduce implementation is Hadoop 1.0.4.

We evaluate our approach by PUMA benchmarks [T+12]. Experimental results show better load balance and shorter task durations achieved by our approach. The performance increase for the job can be as high as 37%.

Contributions of this paper include: 1) We introduce a framework for fine-grained scheduling of MapReduce workload. To the best of our knowledge, this is the first attempt to improve the scheduling of MapReduce workload at the operation level. 2) We model the load balance problem of Reduce operations as a dynamic program, and introduce an algorithm design paradigm to solve it. This paradigm is also applicable to many other problems. 3) We define the sub-problem of selecting Reduce operations for a task slot as the Balanced Subset Sum (BSS) problem. We prove properties of this problem, and give exact and approximation algorithms for it. 4) We design a communication mechanism so that the algorithms above can be transparently incorporated into MapReduce. In addition, we introduce a mechanism named Reduce pipelining to greatly increase resource utilization.

The rest of this paper is organized as follows: Section 2 gives background knowledge. Section 3 gives detailed descriptions of the load balance problem. Section 4 introduces the communication mechanism, which makes it possible to incorporate our algorithms transparently into MapReduce. Our scheduling algorithm is discussed in Section 5. Section 6 gives experimental results. We discuss related work in Section 7 and conclude in Section 8.

## 2 BACKGROUND

An important notation for MapReduce is *job*. It represents all the work that is done after a user submits his computation request and before he gets his result (see Fig. 2). A job contains a number of tasks, and a task contains one or more operations. Hardware resources for MapReduce workload are abstracted as a pool of task slots. A task slot can be either a Map task slot or a Reduce task slot. At any time, each task slot can process at most one task.

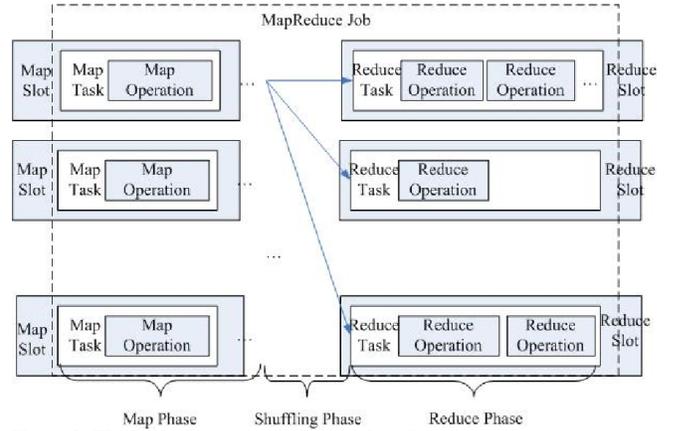

Figure 2. The Basic Structure of a MapReduce Job

Logically, there are 3 major phases in the life cycle of a MapReduce job. The first is the Map phase, whose input is a set of input key-value pairs. These input pairs are split into a number of subsets, and each subset is processed by a Map task and a Map operation. These tasks are distributed across all Map task slots and conducted in parallel. Each Map operation processes its input key-value pairs, and generates some intermediate key-value pairs.

The second phase is the shuffling phase, whose input is the intermediate key-value pairs produced in the Map phase. The shuffling algorithm decides which intermediate pairs are sent to which Reduce tasks. The shuffling algorithm must comply with the constraint that intermediate pairs with the same key are sent to the same Reduce task slot.

The third phase is the Reduce phase. All intermediate pairs with the same key make up the input of a Reduce operation. Similarly, Reduces operations are spread across all Reduce task slots, and processed in parallel. However, Reduce operations must satisfy an additional constraint: all intermediate pairs with the same key are processed by the same Reduce operation on the same Reduce task slot. This is the key difference between Reduce and Map operations. We call it *Reduce Input Constraint*.

Because of the Reduce Input Constraint, intermediate pairs cannot be distributed to Reduce operations at will. This is the fundamental reason for the skew of Reduce operations, and the fundamental reason for the load imbalance of Reduce task slots.

## 3 THE LOAD BALANCE PROBLEM OF MAPREDUCE OPERATIONS

This section discusses the load balance problems for Map and Reduce operations separately. Our objective is to balance the loads of different Map/Reduce task slots. In this study, we assume homogeneous computing cluster, and homogeneous task slots. However, it should be noted that our approach can be easily extended to the heterogeneous cluster, and this will be part of our future work. We assume the numbers of Map and Reduce tasks slots are both equal to $m$. This is consistent with the default settings of standard MapReduce [Wh10].

### 3.1 Load Balance for Map Operations

Since the Map operation does not have any constraints like the Reduce Input Constraint, we can split the set of input key-value pairs at will. Therefore load balance for Map operations is straightforward. Basically, the input key-value pairs can be split into $m$ subsets of equal size, corresponding to $m$ Map operations. Alternatively, the input pairs can be split into $rm$ equal-sized subsets ($r$ = 1, 2, …), corresponding to $rm$ Map operations, so each Map task slot processes $r$ Map operations. Anyway, the total number of input key-value pairs to process is the same for all Map task slots.

$r$ is called the number of *rounds* for Map phase. It has practical impacts. A larger value of $r$ produces smaller operations. This general leads to better load balance, but more initialization and cleaning cost. On the other hand, a smaller value of $r$ leads to worse load balance, but less cost. Therefore, the key of scheduling Map operations is not the scheduling algorithm, but selecting a proper value of $r$ to achieve the optimal trade-off. For standard MapReduce, the default value for $r$ is:

$r$ = input_size / (block_size * $m$)    (3-1)

where block_size refers to the block size of the file system.

### 3.2 Load Balance for Reduce Operations

The scheduling of Reduce operations is more difficult. The fundamental reason lies in the Reduce Input Constraint. Because of this constraint, the set of intermediate key-value pairs cannot be split at will. Instead, the split depends on the number of intermediate pairs with each intermediate key, and we call it the *key distribution of intermediate pairs*. However, this distribution relies on the specific dataset, and cannot be obtained before all Map operations are finished.

*For brevity, for the following discussion, we use pair to refer to intermediate pair, and key to refer to intermediate key.* According to Section 2, all pairs with a certain key are processed by a single Reduce operation, so the number of these pairs decides the load of the Reduce operation. In most cases, the numbers of pairs with different keys are quite different. Therefore the loads of Reduce operations differ greatly. Fig. 1(a) gives an example from our experiments. The largest operation has more than 5000 pairs, while the smallest has less than 10. It is a challenging problem to achieve load balance in the presence of operations with various loads. As we will see shortly, the problem is strongly NP-hard.

Because of this difficulty, current MapReduce adopts simple scheduling strategies. Specifically, the pair $<k, v>$ is sent to the $i$th Reduce task slot [DG04], where

$$i = [\ |\text{Hash}(k)|\ \text{mod}\ m\ ] + 1 \quad (3\text{-}2)$$

*Hash* is a hash function for the key. That means, the Reduce operation corresponding to key $k$ is processed on the $i$th Reduce task slot. It can be verified that this method complies with the Reduce Input Constraint, but the consequent load balance of this method is poor, because the essence of such scheduling strategy is randomly selecting a task slot for each operation. Fig. 1(b) gives an example from our experiment. It can be seen that the load balance produced by standard MapReduce is poor. Another major reason for such poor performance is that current MapReduce does not consider the key distribution of pairs. In other words, they assume all Reduce operations have equal load, which is far from the case (see Fig. 1(a)).

Thus, to achieve load balance, the key distribution of pairs must be taken into account. To formulate this problem, we suppose there are totally $n$ different keys, and the number of pairs with the $j$th key is $k_j$ ($j$ = 1, 2, …, $n$). Equivalently, we suppose there are $n$ Reduce operations, and the number of pairs (load) processed by the $j$th operation is $k_j$.

The schedule for Reduce operations assigns exactly one Reduce task slot to each Reduce operation, and we use a set of binary variables to denote this: $x_{ij}$ ($i$ = 1, 2, …, $m$; $j$ = 1, 2, … $n$). $x_{ij}$ = 1 indicates the $j$th Reduce operation is assigned to the $i$th Reduce task slot, and $x_{ij}$ = 0 otherwise. Therefore, the total number of pairs processed on the $i$th task slot is:

$$p_i = \sum_{j=1}^{n} k_j x_{ij} \quad i = 1, 2 ... m$$

we call it the *load* of the $i$th Reduce task slot. A straightforward evaluation of load balance is the variance of loads for all Reduce task slots:

$$\text{var}(p_1, p_2, ..., p_m) = \frac{1}{m}\sum_{i=1}^{m}(p_i - \bar{p})^2 \text{ where } \bar{p} = \frac{1}{m}\sum_{i=1}^{m} p_i$$

Small values of variances indicate balanced loads. However, this criterion makes the scheduling problem non-linear. Many scheduling problems use the *max-load* instead:

$$msp(p_1, p_2, ..., p_m) = \max(p_1, p_2, ..., p_m)$$

Likewise, small max-load indicates balanced loads. With this criterion, the scheduling problem can be formulated as the following integer program:

min $p$

s.t. $p_i = \sum_{j=1}^{n} k_j x_{ij} \quad i = 1, 2 ... m$

$p_i \leq p \quad i = 1, 2, ..., m$

$\sum_{i=1}^{m} x_{ij} = 1 \quad j = 1, 2, ..., n$

$x_{ij} \in \{0,1\} \quad i = 1, 2, ..., m;\ j = 1, 2, ..., n$

According to the standard notation for scheduling problems [G+79], this problem is denoted as P||C$_{max}$. It has been proved to be strongly NP-hard [Ho98], so there is no polynomial-time exact algorithm for it, unless P = NP. We introduce an algorithm for it in Section 5. To formulate this problem, we need the values of $k_1, k_2, …, k_n$. These values are collected by our communication mechanism, whose details will be described in the next section.

# 4 THE COMMUNICATION MECHANISM

According to the current MapReduce specification, different Map/Reduce operations are totally independent, without any communication. This works well for most scenarios, but for others, it is necessary to provide some communication mechanism to gather the local statistics of each operation to evaluate some global statistics, and then let each operation take action according to such global statistics. According to the previous sections, the problem of scheduling Reduce operations involves collecting key distribution of pairs. So the above mechanism is required for scheduling Reduce operations.

Although the global counter of Hadoop support collecting statistics from all operations and evaluating global information [Wh10], the collected statistics are not aggregated until the end of a MapReduce job, which is useless with respect to making decisions based on it. Besides, the global counter only supports integer data type, but some applications may require collecting other types of data.

In this section, we design a communication mechanism to solve this problem, based on the Master-Slave architecture of MapReduce. According to the MapReduce specification, a daemon is installed on each cluster node. The daemon of the master is named JobTracker, which is responsible for maintaining states of slaves, scheduling Map/Reduce tasks, etc. The daemon of each slave is named a TaskTracker, whose responsibility is to coordinate task slots on the cluster node, and assign tasks to them. Therefore, each operation corresponds to exactly one TaskTracker. According to our communication mechanism, an operation may communicate with its TaskTracker, and a TaskTracker may communicate with the JobTracker. Within a MapReduce job, the communication mechanism works as follows:

1) Each Map operation sends its local statistics to its TaskTracker. The statistics contain pairs of the form: $<key_j, k_j^{(i)}>$, which means the $i$th Map operation (Suppose the ID of this Map operation is $i$) produces $k_j^{(i)}$ pairs with key $key_j$.

2) The TaskTracker receives messages from Map operations on the local host, buffers them, and sends them to the JobTracker.

3) The JobTracker receives messages from TaskTrackers. When the statistics of a job is complete, it aggregates all such information:

$$k_j = \sum_{i=1}^{M} k_j^{(i)} \quad j = 1, 2 ... n$$

$M$ is the total number of Map operations of the job.

4) So far, input of the scheduling problem is complete, so the JobTracker invokes our scheduling algorithm, and sends the resulted schedule to each TaskTracker. The message to each TaskTracker contains a number of pairs of the form: $<key_j, i>$ ($i = 1, 2, …, m$), which means all pairs with key $key_j$ should be sent to the $i$th Reduce task slot.

5) Each TaskTracker forwards the message from the JobTracker to Reduce operations.

6) After receiving message from the TaskTracker, each Reduce operation fetches pairs according to the command of the message.

The above process is transparent to the user of MapReduce. All a user needs is to replace the standard MapReduce library with our extended library. To support customized requirements, we also provide APIs that allows each operation to send customized local statistics, and obtain aggregated global statistics.

*4.1 Reducing the Network Flow*

The mechanism described above effectively supports collecting and aggregating statistics from all operations. However, it may also introduce performance cost. The cost can be divided into two classes. One is the computational overhead for calculating the schedule, and the other is the network overhead for collecting statistics and broadcasting the schedule. According to our experiments in Section 6 (Please see Fig. 8), however, influence of the former is neglectable, so we focus on the network overhead.

In the collecting step, the network flow from each Map operation to its TaskTracker is 8$n$ bytes (we use the *long* type for the number of pairs in each operation, whose width is 8 bytes in Java), so the total network flow from all Map operations to TaskTrackers is 8$Mn$, bytes. The number of TaskTrackers is at most the number of Map operations, so the network flow from all TaskTrackers to the JobTracker is at most 8$Mn$, bytes. In the broadcasting step, the network flow from the JobTracker to each TaskTracker is 4$n$ bytes (we use the *int* type to represent the schedule, whose width is 4 bytes), so is the network flow from each TaskTracker to each Reduce operation. Therefore, the total network flow from the JobTracker to all TaskTrackers is no more than 4$Mn$ bytes. The total network flow from all TaskTrackers to all Reduce operations is 4$Mn$ bytes. In summary, the total network flow in the collecting step is no more than 16$Mn$ bytes, and the total network flow in the broadcasting step is no more than 8$Mn$ bytes.

The analysis above indicates that if the number of Map operations or Reduce operations is large, there will be large network flow. To reduce the network flow, we need to either reduce the number of Map operations, or the number of Reduce operations. The former is easy, since we can split the input pairs at will. For example, this can be accomplished by adjusting the block size of the file system (See formula (3-1)). The latter is difficult, due to the Reduce Input Constraint. To reduce the number of Reduce operations, we combine a set of Reduce operations into an *operation group*. The operation group will be treated as a single unit for scheduling. Suppose the desired number of Reduce operations is $n$, we introduce the following mechanism to combine Reduce operations: operations with keys $key_i$ and $key_j$ are combined, if and only if

$$|Hash(key_i)| \equiv |Hash(key_j)| \pmod{n}$$

It can be verified that after combining, the number of Reduce operation groups is at most $n$.

*4.2 Reduce Pipelining*

Besides the direct overheads discussed above, there is indirect overhead. For standard MapReduce, the destination of a pair can be determined immediately after it is produced.

So the output of a Map operation can be copied to its target Reduce task slot before the end of the Map phase. This copying process can be carried out simultaneously when subsequent Map operations are being processed. This overlap may bring notable performance benefit. However, our approach does not permit this overlap, because the destination of a pair cannot be determined until all Map operations are finished.

To solve the above problem, we introduce another overlap within Reduce tasks, based on the following observation: each Reduce task goes through 3 phases (see Fig. 3(a)): 1) In the copy phase, the pairs produced by Map operations are fetched to the Reduce task slots. 2) In the sort phase, the pairs are sorted by key so that input pairs of the same operation are grouped together. 3) In the run phase, the pairs are processed by invoking the Reduce function. For standard MapReduce, the 3 phases are carried out sequentially. This leads to poor resource utilization, because the 3 phases requires different resources. In general, the copy phase is network I/O intensive, the sort phase is disk I/O intensive, and the run phase is compute-intensive.

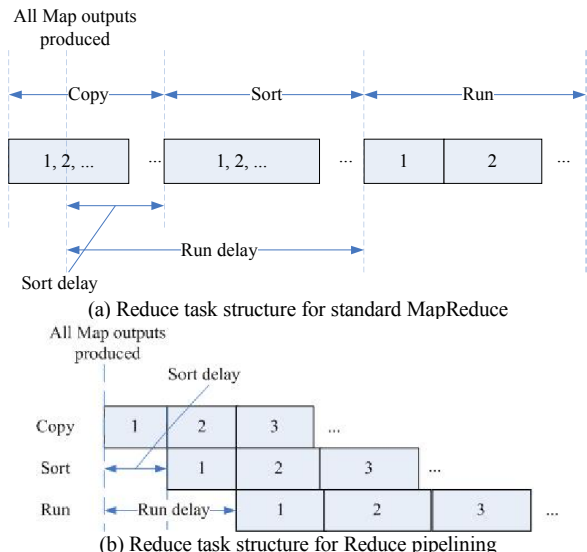

Figure 3. The structure of a Reduce task

To improve resource utilization, we design a pipeline to parallelize the 3 phases (see Fig. 3(b)). The general idea of Reduce pipelining is simple. Suppose a Reduce task has more than one operation (This assumption holds for most Reduce tasks. Please see Section 2 of Supplementary Evaluations). First, the task copies inputs of the first operation, and then sorts and processes them. When sorting the first operation, it simultaneously copies inputs of the second operation. After that, the task sorts the second operation, and simultaneously copies inputs for the third operation and process the first operation, and so on.

Through this mechanism, the input of a Reduce task is split into many small parts, which are processed separately. The split is based on the Reduce operation. That means input of a Reduce operation must be in the same part. According to the property of the Reduce operation, all pairs with the same key are in the same part and processed by the same operation. Therefore, the split does not violate the Reduce Input Constraint, and the correctness of the Reduce task is guaranteed.

Our approach may lead to indirect performance cost, as discussed above. However, Reduce pipelining can cut down the cost. To measure the cost, we define the *sort delay* as the duration from when all Map outputs are produced (when all Map tasks are finished) to when the first Reduce operation enters the sort phase. Similarly, we define the *run delay* as the duration from when all Map outputs are produced (when all Map tasks are finished) to when the first Reduce operation enters the run phase. It should be noted that these definitions also apply to the standard MapReduce. Experimental results for these delays will be given in Section 6.2.2. To make these delays as small as possible, we sort operations by the increasing order of their loads in Reduce pipelining.

## 5 THE KEY DISTRIBUTION BASED SCHEDULING ALGORITHM

This section describes the key distribution based scheduling algorithm. According to Section 3.2, the load balance problem for Reduce operations can be reduced to $P||C_{max}$, which is strongly NP-hard. To solve this problem, we introduce a novel algorithm design paradigm named *dynamic programming decomposition*.

### 5.1 Dynamic Programming Decomposition

As suggested by its name, dynamic programming decomposition is closely related to dynamic programming [KT06]. Dynamic programming works by decomposing the problem into a series of sub-problems, and then constructing solutions to larger and larger sub-problems, until the original problem is solved. To extend one sub-problem's solution to a larger sub-problem's solution, the key is to determine the value of the decision variable. This value is usually obtained by trying all feasible values and choosing the best one.

A wide range of problems can be modeled as dynamic programs, but many of them cannot be solved efficiently by dynamic programming. For these problems, it is inefficient to try all the feasible values for the decision variable, because the number of feasible values is extremely large. So instead of trying every value, we construct and solve an optimization problem to obtain the optimal value of the decision variable. The objective and constraints of the optimization problem are based on properties of the optimal decision variable. What is important is that through this process, the original dynamic programming can be decomposed into a series of sub-problems with lower computational complexity, so we call this method *dynamic programming decomposition*.

We apply dynamic programming decomposition to design scheduling algorithms for the load balance problem. The problem can be modeled by the following dynamic program:

$$msp(S,k) = \max_{U \subseteq S}\left(msp(S-U, k-1), \sum_{j \in U} k_j\right)$$

$$msp(S,1) = \sum_{j \in S} k_j$$

By this model, the schedule is constructed through a series of steps. Each step selects Reduce operations for one Reduce task slot. Set $S$ contains indices for all Reduce operations which are not assigned to any Reduce task slot. $msp(S, k)$ is the optimal max-load for operations in $S$ on $k$ Reduce task slots. $U$ represents the (indices of) Reduce operations selected for the current task slot, and it is the decision variable. Our load balance problem is equivalent to solving $msp(\{1, 2, …, n\}, m)$.

It can be seen that the number of feasible values for $U$ is exponential in the size of $S$, so we set up an optimization problem to determine the optimal $U$. Generally, the total load for the current Reduce task slot should be close to the total load in $S$ divided by $k$. In other words, the total load in $U$ should be as close to the following value as possible:

$$T = \frac{1}{k}\sum_{j \in S} k_j \quad (5\text{-}1)$$

From this discussion the optimization problem for decision variable $U$ can be formulated. We defer the formulation to the next section. We will see that through dynamic programming decomposition, the strongly NP-hard problem is transformed to a series of weakly NP-hard problems.

*5.2 Introducing the BSS problem*

According to the previous section, the key to the scheduling algorithms is to select Reduce operations for one Reduce task slot, so that the total selected load is as close to the *target value T* as possible. This is the optimization problem for determining the decision variable. This problem reminds us of a common problem, the Subset Sum problem. The Subset Sum problem can be formulated as: Given $s$ positive integers $k_1, k_2, …, k_s$, and a target value $T$, try to find a subset of these positive integers whose sum is as large as possible but not greater than $T$ [C+02]. The Subset Sum problem has been proved to be weakly NP-hard.

In practice, however, we found the load balance was poor when adopting the algorithms for Subset Sum to select Reduce operations for Reduce task slots. In particular, task slots whose Reduce operations were assigned last possessed more load, compared with task slots that were assigned first. This is due to the constraint of the Subset Sum problem that the sum of selected integers must be less than or equal to the target value. As far as selecting Reduce operations is concerned, this means the total load of the selected Reduce operations must be less than or equal to $T$. Therefore, in the process of the scheduling algorithm, task slots whose operations were assigned first usually had total load less than $T$, and these missing loads were taken over by other slots whose operations were assigned later. This leaded to poor load balance.

To solve this issue, we design a new model to determine the decision variable of the dynamic program. This problem can be formally stated as: Given $s$ positive integers $k_1, k_2, …, k_s$, and the target value $T$, try to find a subset of these integers whose sum is as close to $T$ as possible. We name it *Balanced Subset Sum* (BSS) problem. It can be proved that BSS is also weakly NP-hard. In practice, when we employed BSS algorithm to select operations for Reduce task slots, the load imbalance issue discussed above almost never existed.

For the rest of this section, we discuss properties of the BSS problem, and its relation to the Subset Sum problem. The BSS problem can be formulated as the following integer program:

$$\min |t - T|$$
$$\text{s.t.} \quad t = \sum_{i=1}^{s} k_i y_i$$
$$y_i \in \{0,1\} \quad i = 1, 2, …s$$

$y_i = 1$ indicates $k_i$ is selected in the subset, and $y_i = 0$ otherwise.

Given positive integers $k_1, k_2, …, k_s$, and target value $T$, a BSS problem instance and a Subset Sum problem instance are defined. For the following discussion, we denote the optimal sum of the BSS instance by BSS($T$). That is, if $y_1^*$ $y_2^*, … y_s^*$ is an optimal solution of the BSS instance, then

$$BSS(T) = \sum_{i=1}^{s} k_i y_i^*$$

Similarly, we denote the optimal sum of the Subset Sum instance by SS($T$). For positive integers $k_1, k_2, …, k_s$, BSS($T$) and SS($T$) can be regarded as functions of the target value $T$. They are both monotone non-decreasing functions.

For some cases, an optimal solution to the BSS instance is also an optimal solution to the Subset Sum instance.

**Lemma 1** Given positive integers $k_1, k_2, …, k_s$, and target value $T$, if $y_1^* y_2^*, … y_s^*$ is an optimal solution of the BSS instance and BSS($T$) $\leq T$, then $y_1^* y_2^*, … y_s^*$ is also an optimal solution for the Subset Sum instance, and BSS($T$) = SS($T$).

**Proof** Suppose $y_1^* y_2^*, … y_s^*$ is not an optimal solution for the Subset Sum instance, there must be some other solution $y_1' y_2', … y_s'$ to the Subset Sum instance, such that

$$\sum_{i=1}^{s} k_i y_i^* < \sum_{i=1}^{s} k_i y_i' \leq T$$

So we have $\left|\sum_{i=1}^{s} k_i y_i^* - T\right| > \left|\sum_{i=1}^{s} k_i y_i' - T\right|$. That means the sum obtained from $y_1' y_2', … y_s'$ is closer to $T$ than that of $y_1^* y_2^*, … y_s^*$, contradicting the assumption that $y_1^* y_2^*, … y_s^*$ is an optimal solution of the BSS instance. So $y_1^* y_2^*, … y_s^*$ must be an optimal solution to the Subset Sum instance, and

$$BSS(T) = SS(T) = \sum_{i=1}^{s} k_i y_i^* \qquad \square.$$

For some BSS problem instances, the value of BSS($T$) may be greater than $T$. The following lemma shows that the former cannot be too much greater than the latter.

**Lemma 2** If $y_1^* y_2^*, … y_s^*$ is an optimal solution to a BSS instance with positive integers $k_1, k_2, …, k_s$ and target value $T$, then BSS($T$) – $k_j < T$ for any $y_j^* = 1$ ($1 \leq j \leq s$).

**Proof** If BSS($T$) ≤ $T$, we are done, so we only consider the case for BSS($T$) > $T$.

For any $y_j^* = 1$, if BSS($T$) – $k_j$ ≥ $T$, we may construct a solution of the BSS instance $y_1', y_2', \ldots y_s'$

$$y_i' = \begin{cases} y_i^* & i \neq j \\ 0 & i = j \end{cases}$$

It can be seen that the sum for this newly created solution is BSS($T$) – $k_j$. The objective value of the new solution is | BSS($T$) – $k_j$ - $T$| = BSS($T$) – $k_j$ – $T$ < BSS($T$) – $T$ = |BSS($T$) – $T$|. So the sum produced by $y_1', y_2', \ldots y_s'$ is closer to $T$ than BSS($T$), contradicting the definition of BSS($T$). Therefore, we must have BSS($T$) – $k_j$ < $T$. □

Having demonstrated these properties, we are ready to explore algorithms for BSS.

*5.3 An Exact Algorithm for the BSS Problem*

According to Lemma 1, the algorithm for Subset Sum can be employed to solve some instances of BSS. However, it fails when BSS($T$) > $T$. Therefore, we give an exact algorithm for BSS in Table 1.

TABLE 1. AN EXACT ALGORITHM FOR THE BSS PROBLEM

|   | Exact_BSS($k_1, k_2, \ldots, k_s, T$) |
|---|---|
| 1 | $L_0 = \{0\}$ |
| 2 | for $i = 1$ to $s$ |
| 3 |     $L'_{i-1} = \{x + k_i \mid x \in L_{i-1}\}$ |
| 4 |     $L_i = L_{i-1} \cup L'_{i-1}$ |
| 5 |     Trim($L_i, T$) |
| 6 | endfor |
| 7 | Find the largest 2 integers $t_1$ and $t_2$ from $L_s$ |
| 8 | if $|t_1-T| < |t_2-T|$ then $t^* = t_1$ |
| 9 | else $t^* = t_2$ endif |
| 10 | Back trace from $t^*$ to get the optimal solution. |

Inputs of the algorithm are operation loads and target value. The algorithm generates $s + 1$ sets of integers $L_0, L_1, \ldots, L_s$. The initial set $L_0$ is generated in line 1, and the remaining sets are generated in the main loop from line 2 to line 6. The $i$th iteration of the loop generates integer set $L_i$. The *Trim* function in line 5 removes some integers from set $L_i$. Specifically, it preserves all integers smaller than $T$, and for integers larger than or equal to $T$, the *Trim* function only preserves the smallest one. By induction on $i$, it can be proved that each integer in $L_i$ is the sum of some subset of $\{k_1, k_2, \ldots, k_i\}$. So when the loop terminates, $L_s$ contains sums of subsets of $\{k_1, k_2, \ldots, k_s\}$.

The largest 2 integers in $L_s$ are found in line 7, and we denote them by $t_1$ and $t_2$. The *if* statement in line 8 and line 9 decides which one of $t_1$ and $t_2$ is closer to $T$, and the one closer to $T$ is denoted by $t^*$. As we shall see, $t^*$ is the optimal sum. The solution of the problem instance is obtained by back tracing integer sets $L_0, L_1, \ldots, L_s$ in line 10. We use a simple example to illustrate the algorithm.

**Example 1** Suppose there are 3 Reduce operations with loads $k_1 = 1, k_2 = 3, k_3 = 2$. They are processed by 2 task slots ($m = 2$). From (5-1), we have the target value $T = 3$.

In the first iteration of the main loop, we have $L'_0 = \{1\}$ and $L_1 = \{0, 1\}$. The Trim function does not change $L_1$.

In the second iteration, we have $L'_1 = \{3, 4\}$ and $L_2 = \{0, 1, 3, 4\}$ before calling *Trim*. After trimming in line 5, we have $L_2 = \{0, 1, 3\}$.

In the third iteration, we have $L'_2 = \{2, 3, 5\}$ and $L_3 = \{0, 1, 2, 3, 5\}$ before trimming. After trimming, $L_3 = \{0, 1, 2, 3\}$.

After the main loop, we have $t_1 = 2$, and $t_2 = 3$. We set $t^* = t_2 = 3$ and by back tracing from $t^*$ we get $y_0 = 1, y_2 = 0, y_3 = 1$, or $y_1 = 0, y_2 = 1, y_3 = 0$, depending on the implementation.

For the sake of efficiency, we implement each integer set $L_i$ ($i = 0, 1, \ldots, s$) as an ordered array. Line 1 of the algorithm takes O(1) time. The time for each iteration of the main loop depends on the sizes of integer sets $L_{i-1}$ and $L_i$. According to the algorithm, each integer set contains distinct nonnegative integers smaller than $T$, and at most one integer greater than or equal to $T$, so $|L_i|$ = O($T$) ($i = 0, 1, \ldots, s$). Hence the total time complexity of the loop is O($sT$). The largest two integers of $L_s$ are the last two, so line 7 takes O(1) time. The *if* statement in line 8 and line 9 also takes O(1) time. The back trace in line 10 takes $s – 1$ steps, and the time for each step depends on the time for finding $t^*$ in $L_i$. This can be done through binary search in time O(log$|L_i|$) = O(log$T$). Therefore, the total time of line 10 is O($s$log$T$), and the total time complexity of Exact_BSS is O($sT$).

The correctness of Exact_BSS is established by the following theorem:

**Theorem 1** The Exact_BSS algorithm correctly obtains the optimal solution of the BSS problem.

**Proof** The key is proving that the value of $t^*$ immediately before back tracing (line 10) is the optimal sum. If this claim is proved, the back trace will correctly obtain the optimal solution.

To prove this claim, let us analyze the main loop. If we eliminate the *Trim* function in line 5 and denote the consequent integer sets by $M_i$ ($i = 0, 1, \ldots, s$), then $M_i$ will contain all the sums that can be obtained from subsets of $\{k_1, k_2, \ldots, k_i\}$, that is

$$M_i = \{s \mid s = \sum_{k \in R} k, R \subseteq \{k_1, k_2, \ldots, k_i\}\}$$

So we have $L_i \subseteq M_i$ ($i = 0, 1, \ldots, s$).

To prove that $t^*$ is the optimal sum, we first prove that the optimal sum is not removed by the *Trim* function. Suppose $y_1, y_2, \ldots y_s$ is an optimal solution to the BSS instance, with optimal sum $t$:

$$t = \sum_{i=1}^{s} k_i y_i$$

It can be seen that $t \in M_s$, and $M_s$ may not be the only set containing $t$. For any $i$, if $t \in M_i$, we must have $t \in M_{i+1}, M_{i+2}, \cdots, M_s$.

Let $p$ be the largest index of selected integer in the optimal solution, i.e. $p = \max\{i \mid y_i = 1\}$. The optimal sum is produced from integers in $\{k_1, k_2, \ldots, k_p\}$, so we have $t \in M_p$. If $t \leq T$, we must have $t \in L_p, L_{p+1}, L_s$. This is because function *Trim* never removes integers smaller than or equal to $T$.

If $t > T$, we must have $t - k_p < T$, according to Lemma 2. By a similar argument, we can prove $t - k_p \in L_{p-1}$. Suppose $t$ is removed from $L_p$ by *Trim*, there must be some other integer $t'$ in $L_p$, so that $T \leq t' < t$. Back tracing from $t'$ will lead to a solution with sum $t'$, which is closer to $T$ than $t$, contradicting our assumption $t$ is the optimal sum. So we must have $t \in L_p$. In a similar manner, we can also prove $t \in L_{p+1}, L_{p+2}, \ldots, L_s$.

Next, we prove the optimal sum must be one of the largest two integers in $L_s$. Suppose $t_1 < t_2$. According to the *Trim* function, at most one of them is greater than or equal to $T$. So we have two cases:

1) $t_1 < T \leq t_2$. For any other integer $t'$ in $L_s$, we have $t' < t_1 < T$. So the distance between $t'$ and $T$ is larger than that of $t_1$ and $T$, and the integer closest to $T$ must be either $t_1$ or $t_2$.

2) $t_1 < t_2 < T$. In this case, all integers in $L_s$ are smaller than $T$, so the integer closest to $T$ is the largest one, namely $t_2$.

In summary, the value of $t^*$ immediately before back tracing is the optimal sum, so the back trace will correctly obtain the optimal solution. □

### 5.4 An Approximation Algorithm for BSS

The Exact_BSS algorithm obtains the optimal solution of BSS. However, it is not a polynomial time algorithm. Its time complexity O($sT$) can be very large in practice, because $T$ is proportional to the total number of pairs of the job. This section discusses how to relax the problem and reduce the time complexity of the algorithm. The basic idea is as follows: The total load assigned to each Reduce task slot need not be equal to $T$, but in a relaxed range around $T$.

To reduce the time complexity of Exact_BSS, a straightforward way is to reduce the number of integers in $L_i$ ($i = 0, 1, \ldots, s$). As far as the job is concerned, a natural solution is to group a number of pairs into a basic unit for scheduling. If each unit contains $\Delta$ pairs, the actual number of scheduling units is reduced by $\Delta$ times. To implement this idea, we insert a step in the Exact_BSS algorithm, and name the new algorithm Relaxed_BSS. Given a BSS instance with positive integers $k_1, k_2, \ldots, k_s$, and target value $T$, the Relax_BSS algorithm first relax each positive integers to the nearest multiple of $\Delta$

$$K_i = \left\lfloor \frac{k_i}{\Delta} + \frac{1}{2} \right\rfloor \cdot \Delta \qquad (i = 1, 2, \ldots, s)$$

The remaining steps of Relax_BSS is identical to those of Exact_BSS, except replacing integers $k_1, k_2, \ldots, k_s$ with the relaxed ones $K_1, K_2, \ldots K_s$.

The time complexity of Relax_BSS is smaller than that of Exact_BSS. To see this, note that for Relax_BSS, each integer in $L_i$ ($i = 0, 1, \ldots, s$) is a distinct multiple of $\Delta$, so the size of these integer sets is O($T/\Delta$), and the time complexity of Relax_BSS is O($sT/\Delta$).

We turn to analysis of the precision of Relax_BSS. We name the BSS instance with relaxed integers $K_1, K_2, \ldots, K_s$ the relaxed BSS instance. According to Theorem 1, Relax_BSS obtains the optimal solution of the relaxed instance. However, the solution is not necessarily the optimal solution to the original BSS instance. Therefore, if the solution obtained by Relax_BSS leads to a sum equal to $T^*$, the sum with respect to the original BSS instance may not be equal to $T^*$, but in a range centered at $T^*$. The following theorem gives the scope of this range.

**Theorem 2** Suppose the sum of the solution obtained by Relax_BSS is $T^*$, the actual sum with respect to the original BSS instance is in the range $[T^* - s\Delta/2, T^* + s\Delta/2)$.

**Proof.** Suppose the solution obtained by Relax_BSS is $y^*_i$ ($i = 1, 2, \ldots, s$). So the sum of this solution is

$$T^* = \sum_{i=1}^{s} K_i y^*_i$$

Suppose the sum of $y^*_i$ ($i = 1, 2, \ldots, s$) with respect to the original BSS instance is $t^*$, that is

$$t^* = \sum_{i=1}^{s} k_i y^*_i$$

According to the rule of relaxing $k_i$ ($i = 1, 2, \ldots, s$), we have
  $K_i - \Delta/2 \leq k_i < K_i + \Delta/2$ ($i = 1, 2, \ldots, s$)
In other words, $-\Delta/2 < K_i - k_i \leq \Delta/2$ for $i = 1, 2, \ldots, s$. Therefore, we have

$$-\frac{\Delta}{2} \sum_{i=1}^{s} y^*_i \leq t^* - T^* = \sum_{i=1}^{s} (k_i - K_i) y^*_i < \frac{\Delta}{2} \sum_{i=1}^{s} y^*_i$$

Because $y^*_i \in \{0, 1\}$ ($i = 1, 2, \ldots, s$), $\Sigma_{1 \leq i \leq s} y^*_i \leq s$. So we have
  $-s\Delta/2 \leq t^* - T^* < s\Delta/2$
  $T^* - s\Delta/2 \leq t^* < T^* + s\Delta/2$. □

We use the following example to illustrate Relax_BSS.

**Example 2** Suppose there are 3 Reduce operations and 2 task slots ($m = 2$). The operation loads are $k_1 = 102$, $k_2 = 304$, $k_3 = 203$. If we group 10 pairs as a single unit ($\Delta = 10$), the relaxed operation loads become $K_1 = 100$, $K_2 = 300$, $K_3 = 200$. By solving this relaxed instance (similar to Example 1), we get the solution $y_1^* = 1$, $y_2^* = 0$, $y_3^* = 1$. The sum of the relaxed instance is $T^* = 100 + 200 = 300$, while the sum of the original BSS instance is $t^* = 102 + 203 = 305$. The difference between $t^*$ and $T^*$ is 5. According to Theorem 2, the difference cannot exceed $3 \times 10/2 = 15$.

To run the Relax_BSS algorithm, the value of $\Delta$ must be determined. If the value is large, the Relax_BSS algorithm runs fast, but precision of the solution is low (according to Theorem 2). On the other hand, a small value of $\Delta$ leads to high precision, but long running time. Therefore, the key is to select proper value of $\Delta$ to achieve the optimal tradeoff. In practice, we use the following value:

$$\Delta_m = \frac{2\eta T}{s} \qquad (5\text{-}2)$$

$\eta$ is a small positive number, which is set by the user to control the *relative error* caused by relaxation. The relative

error is defined as the error incurred by relaxation divided by the target value:

$$rel\_err = |T^* - t^*|/T \quad (5\text{-}3)$$

The reason we use (5-2) to determine the value of $\Delta$ is explained by the following Theorem.

**Theorem 3** If the value of $\Delta$ is set by (5-2), the relative error caused by relaxation is at most $\eta$.

**Proof** According to Theorem 2 and (5-3), the relative error is at most $s\Delta_m/(2T)$. By substituting the value of $\Delta_m$ in (5-2), we have

$$\frac{s\Delta_m}{2T} = \frac{s \cdot 2\eta T}{2sT} = \eta \qquad \square$$

## 6 EXPERIMENTAL EVALUATION

We evaluate our approach by Purdue MapReduce Benchmarks Suite (PUMA) [T+12], and compare it with standard MapReduce (Hadoop 1.0.4, a stable version of Hadoop). In particular, we use the following benchmarks: The Word-Count (WC) benchmark counts the number of occurrences of each word in a set of documents. The Term-Vector (TV) benchmark determines the frequent words in each host. The Inverted-Index (II) benchmark builds a word-to-document index given a set of documents. The Histogram-Movies (HM) benchmark classifies movies based on their ratings. More detailed descriptions of these benchmarks can be found in [T+12]. Each benchmark run on two datasets: the larger one is denoted by L, and the smaller one is denoted by S. So WC_S refers to Word-Count on the smaller dataset, and TV_L refers to Term-Vector on the larger dataset, and so on. Datasets for WC, TV, and II are Wikipedia dump files [Wi13]. Datasets for HM are downloaded from PUMA web site. The dataset sizes are given in Table 2.

TABLE 2. DATASET SIZE (IN GB)

|   | WC   | TV   | II   | HM  |
|---|------|------|------|-----|
| S | 0.94 | 0.94 | 0.94 | 5.1 |
| L | 7.0  | 7.0  | 1.8  | 15  |

The experiments are conduct on a homogeneous cluster with 9 VMs on the IBM RC2 Cloud platform [A+10]. One VM is the master while the other 8 are slaves. Each VM has a virtual CPU of 2.93GHz, 2GB memory and Red Hat 5.5 operating system. According to our test, the bandwidths for network, disk read, and disk write are 14.3MB/s, 45MB/s, and 64MB/s, respectively. Some parameters used in our experiments are given below, while others take default values of Hadoop:

1) The number of Reduce tasks is set to 0.95* <number of VMs> * mapreduce.tasktracker.reduce.tasks.maximum = 0.95 * 8 * 2 ≈ 15, as recommended by Apache Software Foundation [Ma13].

2) For HM, we set the width of intervals to 0.05, because for the default setting, the number of Reduce operations 8 is smaller than the number of Reduce tasks 15.

3) For our approach, if there are more than 120 Reduce operations, we use the mechanism described in Section 4.1 to combine Reduce operations.

4) For our approach, the value of $\eta$ is set to 0.002. According to Theorem 3, the relative error is at most 0.2%.

Before giving results, note that there are 15 Reduce tasks and 16 Reduce task slots (Each VM has 2 Reduce slots, according to the default Hadoop setting). So each Reduce task corresponds to exactly one Reduce task slot, and there is an empty Reduce task slot. Results on this slot will be omitted. Also note that each result is obtained by running 3 jobs and choosing the one with the smallest duration.

### 6.1 Benefits Introduced by Our Approach

Our approach brings two major benefits: better load balance, and shorter task duration.

#### 6.1.1 Better Load Balance

Fig. 4 shows the loads of all Reduce task slots for WC_S. "std" refers to the standard MapReduce (Hadoop 1.0.4), and "impv" refers to the improved MapReduce of our approach. From Fig. 4 it can be seen that the loads are balanced for our approach, due to our key distribution based scheduling algorithm. The load balance is reflected by the max-load. The max-load of standard MapReduce 10488733 is much larger than that produced by our approach 7789497.

Fig. 5 gives max-loads for all cases. It can be seen that for all cases, the max-load produced by our approach is smaller than that produced by standard MapReduce. In Fig. 5, "ideal" refers to the ideal load balance, which is the load when all Reduce task slots share equal load:

$$p_{ideal} = \frac{1}{m}\sum_{i=1}^{n} k_j$$

It can be proved that $p_{ideal}$ is a lower bound of the optimal max-load. It can be seen that for WC, TV and II, the max-load produced by our approach is close to $p_{ideal}$. So the solution produced by our scheduling algorithm is close to the optimal solution. For HM, the max-load produced by our approach is slightly larger than $p_{ideal}$. This is because the Reduce operation loads for HM are highly skewed (see Fig. 1(a)). For example, HM_S has 80 operations, 20 of which have loads larger than 3500. These 20 operations consist of more than 83.4% of the total load. The optimal schedule of the 20 operations can be obtained, with a max-load of 7700, which is larger than $p_{ideal}$ = 6651. So the optimal max-load of the 80 operations is at least 7700, and it can be expected that the schedule produced by our algorithm (with max-load 8651) is close to the optimal schedule. The results for HM_L can be explained similarly.

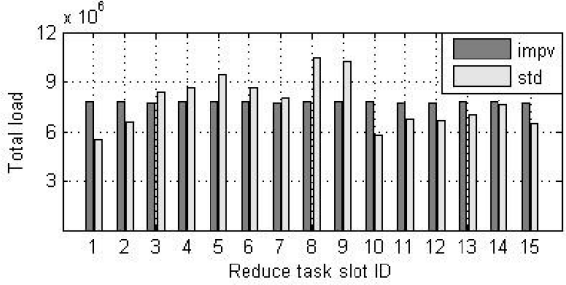

Figure 4. The load of each Reduce task slot for WC_S.

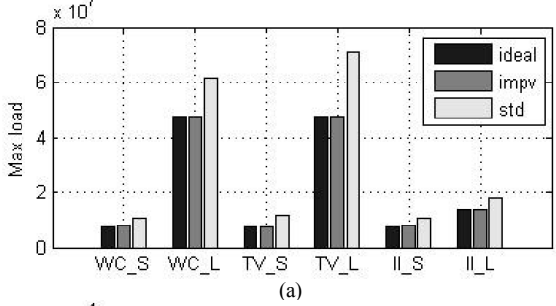
(a)

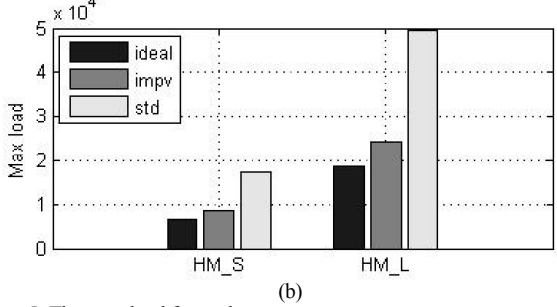
(b)

Figure 5. The max-load for each case.

### 6.1.2 Shorter Task Duration

Fig. 6(a) gives the duration for each Reduce task for TV_S. It can be observed that the overall task duration for our approach is shorter than that of standard MapReduce. This can be explained by two factors: 1) Since our approach produces better load balance, task durations are also balanced. Therefore, the overall time is shorter. 2) Due to Reduce pipelining, 3 phases of the Reduce task is parallelized to some extent, which makes our approach faster.

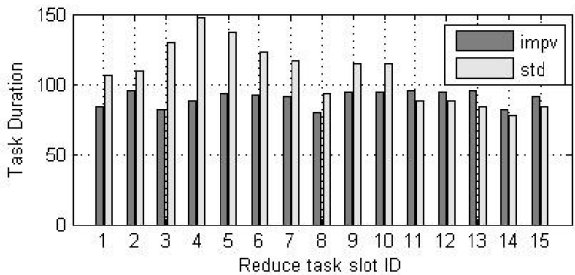
(a) Reduce task durations for TV_S

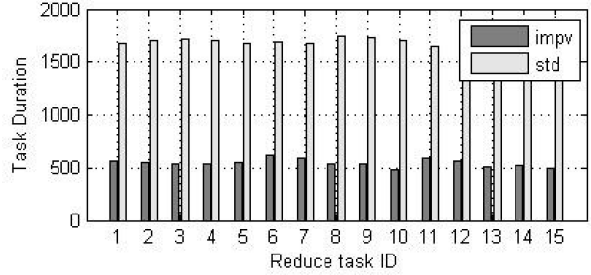
(b) Reduce task durations for TV_L

Figure 6. Reduce task durations for TV (in second)

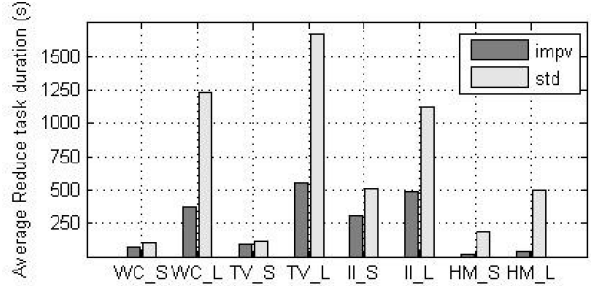

Figure 7. Average Reduce task durations for all cases (in second)

Fig. 6(b) gives the duration for each Reduce task for TV_L. Also, the task duration for our approach is shorter than that of standard MapReduce. However, the difference is greater. In addition to the two factors explained above, this result is explained by an additional factor: 3) Since the input size is large, the number of Map operation is greater than the number of Map task slots (see (3-1)). This results in multiple rounds of Map operations. For standard MapReduce, Reduce tasks start to copy input after the first round of Map operations is over. For our approach, the Reduce tasks cannot begin until all Map operations are finished. So Reduce tasks of standard MapReduce is longer.

Fig. 7 shows the average Reduce task durations. It can be seen that for all cases, the average Reduce durations produced by our approach are smaller than those of standard MapReduce. The results for WC_S, TV_S and II_S are explained by factors 1) and 2) discussed above, because their jobs have only a single round of Map operations. For other cases (WC_L, TV_L, II_L, HM_S, and HM_L), their jobs have multiple rounds of Map operations, so their results are explained by factors 1), 2) and 3).

### 6.2 Costs Introduced by Our Approach

Our approach introduces two types of costs: 1) Compared with the standard MapReduce, our approach involves a centralized scheduling algorithm, which will take some extra time. 2) For our approach, Reduce tasks cannot begin until all Map operations are finished. This may cause some delay compared with standard MapReduce, for which Reduce tasks start when the first round of Map operations are over.

#### 6.2.1 Time Spent on the Scheduling Algorithm

The time spent on our key distribution based scheduling algorithm is shown in Fig. 8. For each case, the time is less than 0.2 second, which is trivial compared with the job duration. In addition, it can be observed that for each benchmark, the time spent for the larger dataset is close to the time spent for the smaller dataset. For example, the time on WC_S is close to that on WC_L. This is explained as follows: although the larger dataset generates more pairs, leading to larger target value *T*, it uses the same η value as used for the smaller dataset. Therefore the value of Δ is also larger (see (5-2)), and the time for Relax_BSS O($sT/\Delta$) remains the same. This indicates that our scheduling algorithm is scalable.

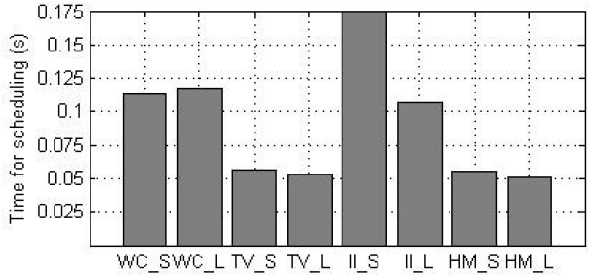
Figure 8. Time spent on the scheduling algorithm (in second)

### 6.2.2 Delays Caused by Our Approach

As discussed in Section 4.2, our approach may introduce delays in Reduce tasks, and these delays can be measured by 2 quantities: sort delay and run delay (see Fig. 3). It should be noted that these delays also exist for standard MapReduce. Fig. 9(a) gives the average sort delay for all Reduce tasks of a job. For WC_S and II_L, the sort delay for our approach is close to that of standard MapReduce. For WC_L, HM_S and HM_L, the sort delays for our approach are larger. This is because the copying phase of standard MapReduce starts much earlier than our approach. For TV_S, TV_L and II_S, sort delays produced by our approach are smaller. This is explained as follows: through Reduce pipelining, the input of Reduce task is divided into a number of parts, each for an operation (group). So the size of each part is smaller than the whole input, and copying the first part by our approach can be finished earlier than copying the whole input by standard MapReduce, even though the latter starts much earlier.

Similar observations and explanations can also be made for run delays, which are given in Fig. 9(b). However, the results for run delays are more favorable for our approach. In particular, the average run delays produced by our approach are smaller for 5 of the 8 cases. This is explained by two factors: 1) Since our approach each time copies a small part of Reduce input, it sorts faster than standard MapReduce, which sorts the whole input. 2) Since our approach each time copies a small fraction, it is more likely that the fraction will be stored and sorted in main memory, which is much faster than storing and sorting it on disk.

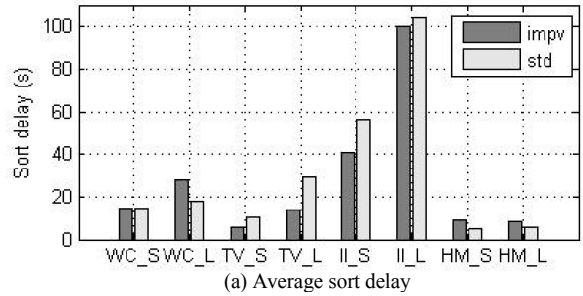
(a) Average sort delay

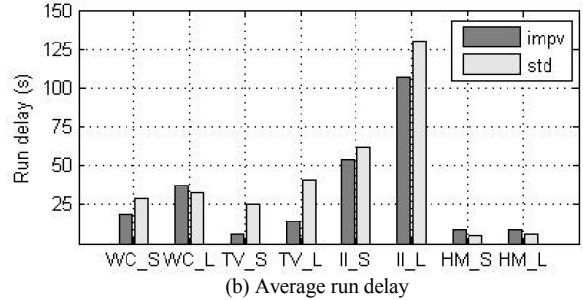
(b) Average run delay

Figure 9. Average delays for all Reduce tasks in a job (in second)

### 6.3 The Overall Effects

All the benefits and costs discussed above are reflected and integrated in the duration of the job, as illustrated by Fig. 10. For all cases, the job duration produced by our approach is smaller than standard MapReduce. This is expectable since from the results above, it can be seen that the cost introduced by our approach is trivial compared with the benefits. To make the distinction clearer, we give the ratio of job duration produced by our approach to the job duration produced by standard MapReduce. From Table 3 it can be seen that the greatest advantage is for HM, whose durations are reduced by 37%. The smallest advantage is for WC_S, whose job duration is reduced by 4.3%.

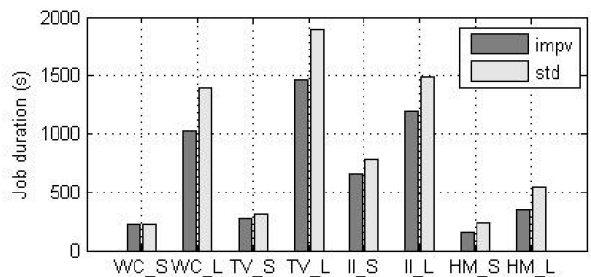
Figure 10. Job duration (in second)

TABLE 3. JOB DURATION RATIOS

|   | WC | TV | II | HM |
|---|---|---|---|---|
| S | 0.9567 | 0.8942 | 0.8389 | 0.6345 |
| L | 0.7339 | 0.7756 | 0.7985 | 0.6314 |

The ratio of job durations produced by our approach to those produced by standard MapReduce

## 7 RELATED WORK

**Scheduling MapReduce Workload at the Job Level.** Tian et al. introduced a dynamic scheduler for MapReduce [T+09]. The scheduler classifies work loads into IO-bound jobs and CPU bound jobs, and schedule them through separate queues. Kang et al. introduced a scheduling scheme for scenarios with multiple computing clusters and multiple MapReduce jobs [K+11]. By batching I/O requests and reducing context switching, the scheme improves performance and enhances fairness between jobs. Sandholm and Lai introduced a dynamic priority scheduler for MapReduce [SL10], which allows the users to dynamically control their allocated capacity. Moseley et al. model MapReduce workload through the classic two-stage flow shop problem, and design a 12-approximation algorithm to minimize the total flow time of multiple jobs [M+11]. A novel framework named SkewReduce is constructed on the basis of MapReduce [K+10]. It transforms a feature-extraction application into a graph of MapReduce jobs, and eliminates skews for these jobs. Although these pieces of work improve the scheduling of MapReduce workload, they are job-level scheduling mechanisms. In other words, they regard a MapReduce job as the basic unit, and try to optimize some criterion in the presence of multiple MapReduce jobs. Therefore, they are totally different from our work. In our approach, the MapReduce operation is the basic unit for scheduling.

**Scheduling MapReduce Workload at the Task Level.** Berlińska and Drozdowski model the MapReduce workload through the Divisible load theory [B+05], and design algorithms to partition and schedule them [BD11]. The focus of their work, however, is on Map tasks, because they assume Reduce tasks have roughly equal execution times. In this study, we find it far from the case and instead focus on scheduling Reduce workload. Zaharia et al.'s algorithm improves the response time by managing speculative tasks [Z+08]. It moves the task backup mechanism [DG04] to the heterogeneous clusters. The SkewTune framework introduced by Kwon et al. is also related to task backup [K+12]. It is an enhancement of the standard task backup. In particular, when a task slot is busy while another task slot is free, the workload of the busy slot is migrated to the free slot. This is similar to standard task backup. However, the backup task and the original task process different parts of the workload, so they complete the workload cooperatively. This is advantageous over standard task backup, since each fraction of the workload is processed only once. Mao et al. introduced a task-level scheduler of MapReduce, which dynamically adjusts the task slots of cluster nodes [M+11b]. Atta design and implement the join algorithm on MapReduce, which effectively deals with the problem of task skew [At10]. However, this result is only applicable to the join algorithm. In summary, these pieces of work improve the scheduling of MapReduce workload at the task level, rather than at the operation level, so they are different from our work.

**Other Improvements of MapReduce**. MapReduce Online improves the standard MapReduce by pipelining Map and Reduce workload [C+10]. For MapRedcue Online, outputs of Map tasks are sent to Reduce tasks immediately after they are produced, without being materialized to local files first. This is the essential difference between MapReduce Online and standard MapReduce. This difference puts Map and Reduce tasks in a pipeline, and can reduce job duration by up to 25% in some scenarios. This mechanism is difference from our Reduce pipelining (see Section 4.2) in the following aspects: 1) For MapReduce Online, the pipeline is between Map and Reduce tasks, while for Reduce pipelining, the pipeline is within the Reduce task. 2) The two stages (Map and Reduce) of MapReduce Online work in sequence. MapReduce Online makes the transition between two stages faster. However, the three stages (copy, sort and run) of Reduce pipelining work in parallel. Therefore, it can be seen that MapRedcue Online is completely different from Reduce pipelining.

Gufler et al. introduced a mechanism to collect and aggregate statistics for MapReduce workload [G+12]. When the size of the statistics is large, it employs algorithms to approximate the global statistics. In particular, it neglects statistics for small operations, and assume they have uniform distribution. This mechanism is different from our approach (see Section 4.1). In particular, although our approach does not have statistics for individual operation, the total load for each task slot is exact, rather than approximated. This is important for our scheduling algorithm to run on exact input.

**Related Scheduling Algorithms.** Sine $P||C_{max}$ is NP-hard, existing algorithms include heuristics and approximation algorithms. Graham introduced a 2-approximation algorithm [Gr66] and a 4/3 approximation algorithm [Gr69] in 1966 and 1969, respectively. These two algorithms are simple and fast, but schedules produced by them are far from the optimal schedule. The approximation schemes introduced by Graham [Gr69] and Sahni [Sa76] may achieve any desired precision, but their time complexities are exponential in the number of task slots ($m$), so they are not applicable to large-scaled clusters. Hochbaum and Shmoys introduced a polynomial time approximation scheme for $P||C_{max}$ [HS87]. This scheme requires a long time to obtain a solution with high precision. For a solution within $1+\varepsilon$ of the optimal max-load, the scheme requires $O((n/\varepsilon)^{1/\varepsilon})$ time. Besides, this scheme is difficult to implement. Our algorithm is fundamentally different from these algorithms. We believe that instead of conflicting these existing algorithms, our algorithm fills the gap between fast but imprecise heuristics, and algorithms that may achieve any desired precision, at the expense of large time complexity.

In Section 3.2, we model the load balance problem as an integer program. Many approximation algorithms or heuristics are designed by integer programming. In general, there are two approaches: One is to design some algorithm for the dual program, and solve the original problem by the solution to the dual program. For example, the polynomial time approximation schemes for $P||C_{max}$ [HS87] and $Q||C_{max}$ [HS88] are both designed through this approach. These schemes are slow, according to our discussion above. The other approach is to relax the integer program to a linear program. Then, the linear program is solved to get a fractional solution. Finally the fractional solution is rounded to an integral solution to the original problem. For example, the 2-approximation algorithm designed by Lenstra et al. adopts this approach [LST90]. This algorithm is fast and easy to implement. However, schedules produced by it are far from optimal schedules. Due to these limitations, we introduce a novel paradigm to solve the scheduling problem: Dynamic Programming Decomposition.

## 8 CONCLUSIONS

MapReduce is one of the most important frameworks in distributed and Cloud computing. An important problem for it is the scheduling of workload. Traditional mechanisms schedule workload at coarse-grained levels, such as job level, or task level. In this study, we schedule MapReduce workload at a fine-grained level: operation level.

The problem of scheduling MapReduce operations, however, is difficult, so current MapReduce implementations adopt simple strategies, leading to poor load balance. In view of this problem, we introduce a set of mechanisms solve it. Specifically, a novel algorithm is introduced which schedules MapReduce operations based on the key distribution of intermediate pairs. To collect this distribution, we design a communication mechanism to extend the MapReduce specification. We define the sub-problem of selecting operations for each Reduce task slot as the Balanced Subset Sum (BSS) problem. We give properties of the BSS problem, and design exact and approximation algorithms for it. The method for designing the scheduling algorithm can be generalized to solve other similar problems. We name it dynamic programming decomposition.

Experiments on PUMA benchmarks show that schedules produced by our algorithm are close to optimal schedules. Therefore, load balance produced by our approach is much better, compared with standard MapReduce. In addition, our algorithm is fast, with running time less than 0.2 second for all cases. Due to our mechanism and algorithm, the job durations are reduced by up to 37%. In the future, we will extend our approach to cases with the heterogeneous cluster and heterogeneous task slots.